%% file: fleming.tex
\documentclass[12pt]{article}
\usepackage{epsfig,graphicx}
\usepackage{fpcp03}

\input fpcp03-macro
\begin{document}

%+ \Chapter{The Large Energy Expansion for B Decays: Soft Collinear Effective Theory}
%+ {}
%+ {Fleming~Sean}

\Title{The Large Energy Expansion for B Decays: Soft Collinear Effective Theory}
\bigskip

%+ \addcontentsline{toc}{chapter}{{\it Fleming~Sean}}
%+ \index{author}{Fleming, S.} 

%%%%%%%%%%%%%%%%%%%%%%%%%%%%%%%%%%%%%
% Label to flag the first page of your contribution
% Replace Perret by your name starting with a capital letter
%
\label{FlemingStart}

%%%%%%%%%%%%%%%%%%%%%%%%%%%%%%%%%%%%%
% Your name
%
\author{ Sean Fleming\index{Fleming, S.} }

%%%%%%%%%%%%%%%%%%%%%%%%%%%%%%%%%%%%%
% Your address
%
\address{Physics Department, Carnegie Mellon University\\
Pittsburgh, PA, 15213 \\
}

\makeauthor\abstracts{
I this talk I give an introduction to the soft collinear effective theory 
by  considering in detail the decay rate for $B \to X_s \gamma$ 
near the endpoint. 
}

\section{Introduction}

Effective field theories provide a simple and elegant method
for calculating processes with several relevant energy
scales
\cite{Weinberg:1978kz,Witten:kx,Georgi:qn,Harvey:ya,Manohar:1995xr,Kaplan:1995uv}.  
Part of the utility of effective theories is that 
they dramatically simplify the summation of logarithms of
ratios of mass scales, which would otherwise make perturbation theory
poorly behaved.  Furthermore the systematic power counting in effective
theories, and the approximate symmetries of the effective field theory
can greatly reduce the complexity of calculations.

Consider as an example the semi-leptonic decay of 
a $\overbar{B}$-meson to a $D$-meson.  
In perturbation theory the one-loop corrections will typically be enhanced by 
$\log(M/\Lambda_{\rm{QCD}})$, where $M$ is a heavy quark mass.  
These logarithms are large enough so that 
$\as \log(M/\Lambda_{\rm{QCD}}) \sim 1$, and the perturbative
expansion  breaks down. Furthermore the nonperturbative physics in 
the decay process is parameterized in terms  of two unknown form factors. 

The power of effective field theories is demonstrated when we consider our 
example within the context of heavy quark effective 
theory (HQET)~\cite{Manohar:dt}. 
In HQET heavy particles have been removed
from QCD so that logarithms in loop integrals are of the form 
$\log(\mu/\Lambda_{\rm{QCD}})$. Furthermore
the complete series of leading logarithms 
$\alpha_s^n \log^n(\mu/\Lambda_{\rm{QCD}})$ is 
straightforward to sum via the renormalization group. The HQET 
Lagrangian has a spin-flavor symmetry which reduces the number of 
form-factors to a single one: the Isgur-Wise function~\cite{Isgur:vq}. In 
addition HQET tells us the normalization of the Isgur-Wise function at a kinematic point. 
Remarkably heavy quark spin symmetry leads to additional relations among 
weak decay form factors. The four form factors which are required to parameterize
matrix elements of vector currents and the four axial vector current form factors
reduce to the single Isgur-Wise function in the heavy quark limit.

\begin{figure}[htbp]
  \epsfxsize=14cm \hfil\epsfbox{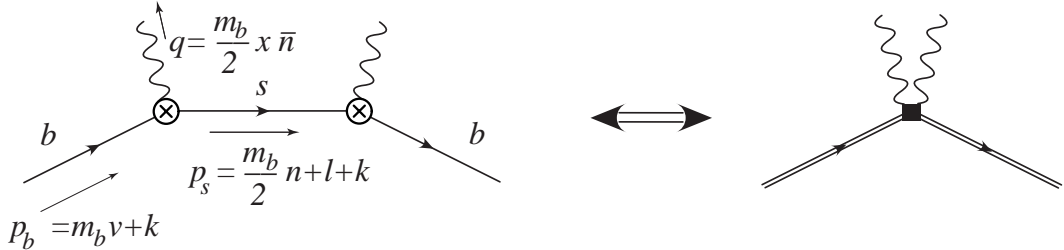}\hfill
     \caption{The OPE for $B\rightarrow X_s\gamma$.}
     \label{opegraphs}
\end{figure}

This example clearly illustrates the power of effective field theories. To motivate  
soft collinear effective theory (SCET) I will consider another example: the decay rate for 
$\overbar{B} \to X_s \gamma$.  The dominant contribution to this decay 
arises from the magnetic penguin operator~\cite{Grinstein:tj}
\begin{equation}
\hat{O}_7 = {e \over 16 \pi^2} m_b\; \bar{s} \, \sigma^{\mu \nu}
{1 \over 2} (1+ \gamma_5) b \; F_{\mu \nu} \,,
\label{o7}
\end{equation}
where the strange quark mass has been set to zero.
The operator product expansion (OPE) for this decay is illustrated in
Fig.\ \ref{opegraphs}.  We write the momenta of the $b$ quark, photon,
and light $s$ quark jet as
\begin{equation}\label{momenta}
\label{kinematics}
p_b^\mu=m_b v^\mu+k^\mu,\ \ q^\mu={m_b\over 2} x \bar n^\mu,\ \ 
p_s^\mu={m_b\over 2} n^\mu + l^\mu + k^\mu
\end{equation}
where, in the rest frame of the $B$ meson,
\begin{equation}
v^\mu=(1,\vec 0),\ \ n^\mu=(1,0,0,-1),\ \ \bar
n^\mu=(1,0,0,1). 
\end{equation}
Here $k^\mu$ is a residual momentum of order $\Lambda_{\rm QCD}$, and
$l^\mu={m_b\over 2}(1-x) \bar n^\mu$, where $x = 2 E_\gamma/m_b$.  The
invariant mass of the light $s$-quark jet
\begin{equation}\label{invmass}
p^2_s \approx m_b \, n \cdot (l+k) = m^2_b (1-x+\hat k_+)\,,
\end{equation}
(where $\hat k_+ = k_+/m_b$) is $O(m^2_b)$ except near the endpoint of
the photon energy spectrum where $x \to 1$. Inclusive quantities are
calculated via the OPE 
by taking the imaginary part of the graphs on the 
left hand side of the double arrow in
Fig.\ \ref{opegraphs} and expanding in powers of $k^\mu/\sqrt{p^2_s}$.
As long as $x$ is not too close to the endpoint, this is an expansion
in powers in $k^\mu/m_b$, which matches onto local operators
shown graphically on the right hand side of the double arrow in Fig.\ \ref{opegraphs}.
This leads to an expansion for the photon energy spectrum as a
function of $x$ in powers of $\alpha_s$ and $1/m_b$~\cite{Falk:1993dh,Kapustin:1995nr}:
\begin{eqnarray}\label{qcdinclusive}
\frac{d\Gamma}{dx} &=& \Gamma_0(\mu) 
\left[ \frac{m_b(\mu)}{m_b} \right]^2  \left\{  \left[1-\frac{\alpha_s C_F}{4
\pi} \left( 2 \log \frac{\mu^2}{m_b^2} +5 + \frac{4}{3} \pi^2 \right)
\right] \delta(1-x) \right.  \nonumber\\ && \;\;\;\;\;\;\;\;\;\left. +
\frac{\alpha_s C_F}{4\pi} \left[ 7 + x -2 x^2 - 2(1+x)\log(1-x) -
\left( 4 \frac{\log(1-x)}{1-x} + \frac{7}{1-x} \right)_+ \right]
\right. \nonumber \\
&& \;\;\;\;\;\;\;\;\ \left.+{1\over 
2m_b^2}\left[(\lambda_1-9\lambda_2)\delta(1-x)
-(\lambda_1+3\lambda_2)\delta^\prime(1-x)-{\lambda_1\over 
3}\delta^{''}(1-x)\right]\right\}
\nonumber \\
&& \;\;\;\;\;\;\;\;\;\;+O(\alpha_s^2, 1/m_b^3),
\label{fullrate}
\end{eqnarray}
where
\begin{equation}
\Gamma_0(\mu)  = {G^2_F |V_{tb} V^*_{ts}|^2 \alpha |C_7(\mu)|^2 \over 32
\pi^4} m^5_b   \,,
\label{gamma0}
\end{equation}
and the subscript ``+" denotes the usual plus distribution.
The parameters $\lambda_1$ and $\lambda_2$ are matrix elements of
local dimension five operators.

Near the endpoint of the photon spectrum, $x\rightarrow 1$, both the
perturbative and nonperturbative corrections are singular and the OPE
breaks down.  The severity of the breakdown is most easily seen by
integrating the spectrum over a region $1-\Delta < x < 1$.  When
$\Delta\leq \lqcd/m_b$ the most singular terms in the $1/m_b$
expansion sum up into a nonperturbative shape function of
characteristic width $\lqcd/m_b$\cite{Neubert:1993ch,Bigi:1993ex}.  
The perturbative series is of the form
\begin{equation}
{1\over\Gamma_0}\int_{1-\Delta}^1 {d\Gamma\over dx}=1+{\alpha_s C_F\over 4
\pi}\left(-2
\log^2\Delta-7\log\Delta+\dots\right) + O \left( \alpha_s^2 \right),
\end{equation}
where the ellipses denote terms that are finite as $\Delta\rightarrow
0$.  These Sudakov logarithms are large for $\Delta\ll 1$, and can
spoil the convergence of perturbation theory. The full series has been
shown to exponentiate \cite{Korchemsky:1994jb,Akhoury:1995fp}, which
sums the leading and next-to-leading logarithms.

{}In general, ``phase space'' logarithms are to be expected whenever a
decay depends on several distinct scales.  For example,
in $b\rightarrow X_c e\bar\nu_e$
decay the rate calculated with the OPE performed at $\mu=m_b$ contains logarithms
of $m_c/m_b$, which become large in the $m_b \gg m_c$ limit.  
In \cite{Bauer:1996ma} an effective theory was used to 
run from $m_b$ to $m_c$, summing phase space logarithms of the ratio $m_c/m_b$.    Similarly,
in $b\rightarrow X_s\gamma$
near the endpoint of the photon energy spectrum the invariant mass of
the light quark jet scales as $m_b\sqrt{1-x}$, and is widely separated
from the scale $\mu=m_b$ where the OPE is performed.  In order to sum
logarithms of $\Delta$ (or the more complicated plus distributions in
the differential spectrum, Eq.\ (\ref{qcdinclusive})) we would expect
to have to switch to a new effective theory at $\mu=m_b$, use the
renormalization group to run down to a scale of order $m_b\sqrt{1-x}$,
at which point the OPE is performed.  In fact, we will see that
the situation is more complicated than this. 

We are then left with the question of the appropriate theory below the
scale $m_b$.  To see where we might begin let us return to the expressions for the
$s$-quark momentum and the invariant mass given in Eq.~(\ref{momenta}) and  
Eq.~(\ref{invmass}) respectively. For $1-x \sim {\cal O} (\Lambda_{ {\rm QCD}}/ m_b)$ we find 
$l^\mu \sim k^\mu \sim \Lambda_{{\rm QCD}}$, and the invariant mass of the $s$-quark
is ${\cal O}(m_b  \Lambda_{{\rm QCD}})$. On the other hand the momentum of
the $s$-quark has a large component of order $m_b$ in the light-cone direction $n^\mu$
with residual momentum of order $\Lambda_{{\rm QCD}}$. Thus in this kinematic
region the $s$-quark is light-like.  Given that the form of the light-like momentum,
$p^\mu_s = (m_b/2) n^\mu + \tilde{k}^\mu$ with $\tilde{k}^\mu \sim \Lambda_{{\rm QCD}}$, 
is the same as the form of the heavy quark momentum, $p^\mu_b = m_b v^\mu + k^\mu$,
with the time like vector $v^\mu$ replaced with the light-like vector $n^\mu$ it is very tempting to 
introduce an effective theory of light-like Wilson lines, much as HQET is an 
effective theory of time-like Wilson lines\cite{Grozin:1994ni}. Such an effective theory, christened 
the large energy effective theory (LEET), was proposed many years ago by Dugan and
Grinstein\cite{Dugan:1990de}.  However, a simple attempt at matching shows that LEET does not 
reproduce the infrared physics of QCD~\cite{Bauer:2000ew}. 
The problem is that LEET only describes the coupling of light-like
particles to soft gluons, but does not describe the splitting of an
energetic particle into two almost collinear particles.  

\section{Soft Collinear Effective Theory}

In the rest frame of the heavy hadron the light
particles in the decay move close to the light cone direction $n^\mu$ and their dynamics is
best described in terms of light cone coordinates $p = (p^+, p^-, p_\perp)$,
where $p^+ = n \cdot p$, $p^- = \bar{n} \cdot p$.  For large
energies the different light cone components are widely separated, with $p^-
\sim Q$ being large, while $p_\perp$ and $p^+$ are small. Taking the small
parameter to be $\lambda \sim p_\perp/p^-$ we have
\begin{equation} \label{scaling}
 p^\mu = \bar{n} \cdot p\, \frac{n^\mu}{2} + p_\perp^\mu + 
   n\cdot p\,\frac{\bar{n}^\mu}{2} = {\cal O}(\lambda^0) +{\cal O}(\lambda^1) +
   {\cal O}(\lambda^2) \,,
\end{equation}
where we have used $p^+p^- \sim p_\perp^2 \sim Q^2 \lambda^2$ for fluctuations near
the mass shell.  Thus the light-cone momentum components of collinear particles
scale like $k_c = Q(\lambda^2, 1,\lambda)$. The collinear quark can emit either 
a gluon collinear to the large momentum direction or a gluon with momentum 
scaling $k_{us} = Q(\lambda^2,\lambda^2,\lambda^2)$ (referred to as an ultra-soft
or usoft gluon). For scales above the typical off-shellness of the collinear
degrees of freedom, $k_c^2 \sim (Q \lambda)^2$, both gluon modes
are required to correctly reproduce all the infrared physics of the full
theory. This was described in \cite{Bauer:2000ew}, where it was shown that at a scale
$\mu\sim Q$, QCD can be matched onto an effective theory that contains heavy
quarks and light collinear quarks, as well as usoft and collinear gluons.

The SCET Lagrangian can be obtained at tree level by expanding
the full theory Lagrangian in powers 
of $\lambda$~\cite{Bauer:2000yr}. We start from the
QCD Lagrangian for massless quarks and gluons
\begin{eqnarray} \label{QCD}
 {\cal L}_{\rm QCD} &=& \bar \psi\: i \Dslash\: \psi  \,,
\end{eqnarray}
where the covariant derivative is $D_\mu = \partial_\mu - i g T^a A^a_\mu$.  
We begin by removing the large
momenta from the effective theory fields, similar to the construction of
HQET. In HQET there are two relevant momentum scales, the mass of
the heavy quark $m$ and $\Lambda_{\rm QCD}$. The scale $m$ is separated from
$\Lambda_{\rm QCD}$ by writing $p=m v+k$, where $v^2 = 1$ and the residual
momentum $k\ll m$.  The variable $v$ becomes a label on the effective
theory fields.  Our case is slightly more complicated because there are three
scales to consider. We split the momenta $p$ by taking
\begin{eqnarray}
  p = \tilde p + k \,, \qquad\mbox{where}\quad
  \tilde p\equiv \frac12 (\bar{n}\cdot p)n + p_\perp \,.
\end{eqnarray}
The ``large'' parts of the quark momentum $\bar{n}\cdot p\sim 1$ and $p_\perp\sim
\lambda$, denoted by $\tilde p$, become a label on the effective theory field,
while the residual momentum $k^\mu \sim \lambda^2$ is dynamical. 

The large momenta $\tilde p$ are removed by defining a new field
$\psi_{n,p}$ through
\begin{eqnarray}\label{xidef}
 \psi(x) = \sum_{\tilde p} e^{-i\tilde p\cdot x} \psi_{n,p}\,.
\end{eqnarray}
A label $p$ is given to the $\psi_{n,p}$ field, with the understanding that only
the components $\bar{n}\cdot p$ and $p_\perp$ are labels. Derivatives
$\partial^\mu$ on the field $\psi_{n,p}(x)$ give order $\lambda^2$
contributions. For a particle moving along the $n^\mu$ direction, the four
component field $\psi_{n,p}$ has two large components $\xi_{n,p}$ and two small
components $\xi_{\bar{n},p}$. These components can be obtained from the field
$\psi_{n,p}$ using projection operators
\begin{eqnarray}
 \xi_{n,p} = \frac{{n\!\!\!\slash} {\bar n\!\!\!\slash}}{4}\: \psi_{n,p}\,,\qquad
 \xi_{\bar{n},p} = \frac{{\bar n\!\!\!\slash} {n\!\!\!\slash}}{4}\: \psi_{n,p}\,,
\end{eqnarray}
and satisfy the relations
\begin{eqnarray}
& &\frac{{n\!\!\!\slash} {\bar n\!\!\!\slash}}{4}\: \xi_{n,p} = \xi_{n,p}\,, \quad {n\!\!\!\slash}\, 
  \xi_{n,p}=0\,,  \nonumber \\
& &\frac{{\bar n\!\!\!\slash} {n\!\!\!\slash}}{4}\: \xi_{\bar{n},p} = \xi_{\bar{n},p}\,, \quad 
  {\bar n\!\!\!\slash}\, \xi_{\bar{n},p}=0 \,.
\end{eqnarray}
In terms of these fields the quark part of the QCD Lagrangian given in
Eq.~(\ref{QCD}) becomes
\begin{eqnarray}\label{L_split}
 {\cal L} &=& \sum_{\tilde p, \tilde p'}\bigg[ \bar\xi_{n,p'} \frac{{\bar n\!\!\!\slash}}{2} 
 \Big( in\cdot D \Big) \xi_{n,p} 
 + \bar\xi_{\bar{n},p'} \frac{{n\!\!\!\slash}}{2}\Big( \bar{n}\cdot p +  i\bar{n}\cdot D\Big) 
   \xi_{\bar{n},p} \nonumber\\[4pt]
&&\qquad +\bar\xi_{n,p'} \Big({p \!\!\!\slash}_\perp + i {D\!\!\!\!\slash}_\perp \Big) 
 \xi_{\bar{n},p} 
 + \bar\xi_{\bar{n},p'} \Big({p \!\!\!\slash}_\perp + i {D\!\!\!\!\slash}_\perp \Big) \xi_{n,p}
 \bigg] \,.
\end{eqnarray}
Since the derivatives on the fermionic fields yield momenta of order $k \sim
\lambda^2$ they are suppressed relative to the labels $\bar{n}\cdot p$ and
$p_\perp$. Without the $\bar{n}\cdot D$ and $D_\perp$ derivatives, $\xi_{\bar{n},p}$ is
not a dynamical field. Thus, we can eliminate $\xi_{\bar{n},p}$ at tree level by
using the equation of motion
\begin{eqnarray} \label{frdefn}
  (\bar{n}\cdot p + \bar{n}\cdot iD) \xi_{\bar{n},p} = ({p\!\!\!\slash}_\perp +
i{D\!\!\!\!\slash}_\perp ) \frac{{\bar n\!\!\!\slash}}{2} \xi_{n,p} \,.
\end{eqnarray}
Eqs.~(\ref{L_split}) and (\ref{frdefn}) result in a Lagrangian
involving only the two components $\xi_{n,p}$:
\begin{eqnarray}  \label{preLc}
 {\cal L} &=& \sum_{\tilde p, \tilde p'} e^{-i(\tilde p - \tilde p')\cdot x}
  \bar\xi_{n,p'} \left[ n\cdot iD +({p\!\!\!\slash}_\perp + i{D\!\!\!\!\slash}_{\perp} )
  \frac{1}{\bar{n}\cdot p + \bar n \cdot iD} ({p\!\!\!\slash}_\perp +i {D\!\!\!\!\slash}_{\perp} )
  \right] \frac{{\bar n\!\!\!\slash}}{2} \xi_{n,p} \,.
\end{eqnarray}
Here the summation extends over all distinct copies of the fields labelled by
$\tilde p, \tilde p'$.  The gluon field in $D^\mu$ includes collinear
and usoft parts, $A^\mu \to A^\mu_c + A^\mu_{us}$. The two types of gluons are
distinguished by the length scales over which they fluctuate.  Fluctuations of
the collinear gluon fields $A^\mu_c$ are characterized by the scale $q^2\sim
\lambda^2$, while fluctuations of the usoft gluon field $A^\mu_{us}$ are
characterized by $k^2\sim \lambda^4$. Since the collinear gluon field has large
momentum components $\tilde q\equiv (\bar{n}\cdot q,q_\perp)$, derivatives acting on
these fields can still give order $\lambda^{0,1}$ contributions. To make this
explicit we label the collinear gluon field by its large momentum components
$\tilde q$, and extract the phase factor containing $\tilde q$ by redefining the
field: $A_c(x) \to e^{-i\tilde q \cdot x} A_{n,q}(x)$.  Inserting this into
Eq.~(\ref{preLc}) one finds
\begin{eqnarray} \label{pre2Lc}
  {\cal L} &=& \sum_{\tilde p, \tilde p', \tilde q} e^{-i(\tilde p-\tilde
  p')\cdot x} \bar\xi_{n,p'} \Bigg[ n\cdot iD\, +g e^{-i\tilde q\cdot x} n\cdot
  A_{n,q} + \Big( {p\!\!\!\slash}_\perp + i{D\!\!\!\!\slash}_{\perp} + ge^{-i\tilde q\cdot x}
  {A\!\!\!\slash}_{n,q}^\perp\Big) \\ 
  & & \qquad \times \frac{1}{\bar{n}\cdot p + \bar{n}\cdot
  iD + ge^{-i\tilde q\cdot x} \bar{n}\cdot A_{n,q}} \Big({p\!\!\!\slash}_\perp +i
  {D\!\!\!\!\slash}_{\perp}+ ge^{-i\tilde q\cdot x} {A\!\!\!\slash}_{n,q}^\perp \Big) \Bigg]
  \frac{{\bar n\!\!\!\slash}}{2} \xi_{n,p} \,.\nonumber
\end{eqnarray} 
The covariant derivative is defined to only involve usoft
gluons. 
We immediately notice a problem: derivatives in the 
the last term in brackets above can act on the phase 
factors associated with the collinear gluon fields. Thus these derivatives
can result in terms of order $\lambda^0,\lambda^1$, while we  want
all derivatives to scale as $\lambda^2$.  

\begin{table}[t!]
\begin{center}
\begin{tabular}{|ccccccc|}
\hline \\
            & heavy quark & collinear quark\ & usoft gluon\ & & collinear gluons &
  \\\hline 
  Field & $h_v$ & $\xi_{n,p}$ & $A_{us}^\mu$ & $\bar n\cdot A_{n,q}$ 
            & $n\cdot A_{n,q}$ & $A_{n,q}^\perp$ \\ 
  Scaling & $\lambda^3$ & $\lambda$ & $\lambda^2$ & $\lambda^0$ 
            & $\lambda^2$ & $\lambda$  \\
\hline 
\end{tabular}
\end{center}
{\caption{Power counting for SCET fields.}
\label{table_pc} }
\end{table}

We need to split up the derivative into a piece that 
acts on  the large components of the collinear momentum,  and a residual 
piece that is ${\cal O}(\lambda^2)$. Towards this end we introduce a projection
operator ${\cal P}$ which only acts on the large label of the collinear 
fields~\cite{Bauer:2001ct}. For any function $f$
\begin{eqnarray}
&& f(\bar{{\cal P}}) \phi^\dagger_{q_1} \cdots \phi^\dagger_{q_m} 
\phi_{p_1} \cdots \phi_{p_n} 
\nonumber \\
&& \hspace{3ex} f(\bar{n} \cdot p_1 + \dots + \bar{n} \cdot p_n - 
\bar{n} \cdot q_1 - \dots \bar{n} - \cdot q_m)
\phi^\dagger_{q_1} \cdots \phi^\dagger_{q_m} 
\phi_{p_1} \cdots \phi_{p_n} \,,
\end{eqnarray}
where $\bar{{\cal P}} \equiv \bar{n} \cdot {\cal P}$. Then we have
\begin{equation}
i \partial^\mu e^{- i p \cdot x} \phi_{n,p}(x) = 
e^{- i p \cdot x}({\cal P}^\mu + i \partial^\mu )  \phi_{n,p}(x) \,,
\end{equation}
and the phases involving large momentum components can be removed 
from the Lagrangian in Eq.~\ref{pre2Lc} as long as we adopt a convention 
that there is an implicit sum over labels, and that total  label momentum is
conserved. 

Finally, we expand Eq.~(\ref{pre2Lc}) in powers of $\lambda$.  To simplify the
power counting we follow the procedure of moving all the dependence 
on $\lambda$ into the interaction terms of the action to make the
kinetic terms of order $\lambda^0$~\cite{Luke:1996hj,Grinstein:1997gv,Luke:1997ys}. 
This is done by assigning a $\lambda$
scaling to the effective theory fields as given in Table~\ref{table_pc}. The
power counting in Table~\ref{table_pc} gives an order one kinetic term for
collinear gluons in an arbitrary gauge.  In generalized covariant gauge
\begin{eqnarray}
 \int d^4x\: e^{ik\cdot x}\: \langle 0 |\: T\: A_c^\mu(x) A_c^\nu(0)\: 
 | 0\rangle =\frac{-i}{k^2} \Big( g^{\mu\nu} - \alpha \frac{k^\mu
 k^\nu}{k^2} \Big)
\end{eqnarray}
and the scaling of the components on the right and left hand side of this
equation agree. Note that $x$ must be rescaled as 
well: $(x^+, x^-,x^\perp)  \to ( x^+ / \lambda^2 , x^- , x^\perp / \lambda)$. 
With  this power counting all interactions scale as $\lambda^n$ with $n \geq 0$, and
the leading SCET Lagrangian for the collinear quark sector is 
\begin{equation}\label{scetlag}
{\cal  L}^{(0)}_{\xi \xi} = \bar{\xi}_{n,p'} \bigg\{ i n\cdot D + i {D\!\!\!\!\slash}^\perp_c
\frac{1}{i\bar{n} \cdot D_c} i   {D\!\!\!\!\slash}^\perp_c \bigg\} \frac{{\bar n\!\!\!\slash}}{2}
\xi_{n,p}\,,
\end{equation}
where $i n\cdot D = i n \cdot \partial + g n\cdot A_{n,q} + g n\cdot A_{us}$,  
$i\bar{n} \cdot D_c = \bar{{\cal P}} + g \bar{n}\cdot A_{n,q}$, and 
$i D^{\perp \mu}_c = {\cal P}^{\perp\mu} + g A^{\perp\mu}_{n,q}$. 

As I mentioned at the beginning of the talk an important aspect of effective field
theories is the approximate symmetries that are manifest in the leading order
Lagrangian. The SCET Lagrangian presented above has a global helicity spin 
symmetry, which for example can lead to a reduction in the number of form 
factors needed to parameterize heavy-to-light decays. In addition 
SCET has a powerful set of gauge symmetries~\cite{Bauer:2001yt}. 
Specifically the collinear and usoft fields each have their own gauge transformation that leave the 
Lagrangian invariant. Collinear gauge transformations are the subset of QCD
gauge transformations where $\partial^\mu U(x) \sim Q (\lambda^2, 1 , \lambda)$,
and usoft gauge transformations are those where $\partial^\mu V(x) \sim  Q \lambda^2$. 
The invariance under each of these transformations is a manifestation of 
scales of order $Q$ or greater having been removed from the theory, since 
any gauge transformation that would change a usoft gluon into a collinear gluon 
would imply a boost of order $Q$. The gauge transformations for the SCET fields are
shown in Table~\ref{gauge}. The usoft field acts like a classical background
field in which the collinear particle propagates, and the collinear fields transform similarly 
to a global color-rotation under a usoft gauge transformation. 
In a moment we will see why the gauge invariance structure is so powerful. But first I 
want to comment on something called reparameterization invariance.

\begin{table}[t!]
\begin{center}
\begin{tabular}{|cccccc|}
 \hline
  & \hspace{0.6cm}Fields\hspace{0.6cm}  & Collinear $U$  & Usoft $V$ & \\ 
  \hline 
  & $\xi_{n,p}$ & $U \ \xi_{n,p}$ &  $V \,\xi_{n,p}$ \\
  & $A_{n,q}^\mu$ & \hspace{0.2cm}$U \: A^\mu_{n,q}\: U^\dagger + 
      \frac{1}{g}\, U \Big[i\cD^\mu \: U^\dagger \Big]$ 
      & $V \, A_{n,q}^\mu\, V^\dagger$ \\ 
 & & & & \\
  & $q_{us}$ & $q_{us}$ & $V \, q_{us}$  & \\
  & $A_{us}^\mu$ & $A_{us}^\mu$ & $V\Big( A_{us}^\mu +
  \frac{i}{g} \partial^\mu \Big) V^\dagger$  & \\ 
  \hline
\end{tabular}
\end{center}
{\caption{Gauge transformations for the collinear and usoft fields. Here 
$i {\cal D}^\mu  \equiv 
(n^\mu / 2) \bar{{\cal P}} +{\cal P}^\mu_\perp + (\bar{n}^\mu / 2) i n\cdot D$.}
\label{gauge} }
\end{table}

Reparametrization invariance (RPI)~\cite{Chay:2002vy,Manohar:2002fd} in SCET 
is a manifestation of the Lorentz symmetry 
that was broken by introducing the vectors $n$ and $\bar{n}$. Lorentz symmetry tells
us that any choice of light-like vectors $n$ and $\bar{n}$ is equally good as long as 
$n^2 = 0$, $\bar{n}^2 = 0$, and $n\cdot\bar{n} = 2$. This implies that SCET 
operators must be invariant under the most general set of transformations that satisfy the 
conditions just enumerated. These transformations fall into three 
catagories: I) $n \to n +\Delta_\perp$, $\bar{n} \to \bar{n}$, 
II) $n \to n$, $\bar{n} \to \bar{n} + \epsilon_\perp$, and
III) $n \to e^\alpha n$, $\bar{n} \to e^{-\alpha} \bar{n}$, 
where $\Delta_\perp \sim \lambda$ and
$\epsilon_\perp, \alpha \sim \lambda^0$. Requiring invariance of operators under these 
transformations results in powerful constraints on the forms of operators. In fact RPI
is an essential tool for deducing subleading corrections in SCET~\cite{Pirjol:2002km}. 

\section{Heavy-to-Light current}

\begin{figure}[!t]
 \centerline{\mbox{\epsfysize=4.0truecm \hbox{\epsfbox{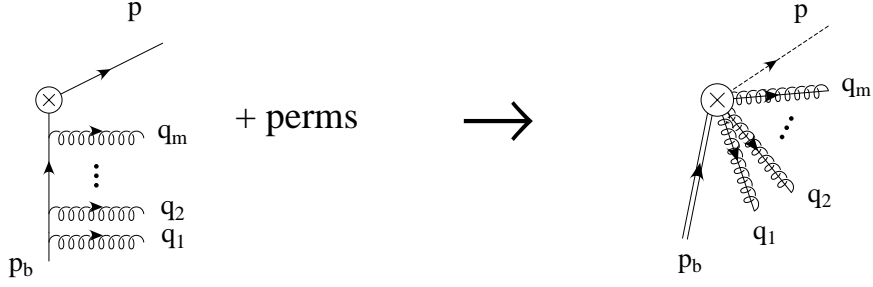}}  }}
{\caption[1]{Matching for the order $\lambda^0$ Feynman rule for the
heavy to light current with $n$ collinear gluons. All permutations of crossed 
gluon lines are included on the left.} 
 \label{fig_current} }  
\end{figure}

Next I discuss matching the SCET heavy to light currents. 
To perform the matching, first consider the simpler case of an Abelian gauge
group. In this case calculating the full theory graph with $m$ gluons in
Fig.~\ref{fig_current}, expanding in powers of $\lambda$, and putting the result
over a common denominator gives
\begin{eqnarray} \label{cm}
  c_m(\mu=m_b) = \frac{1}{m!}\: \prod_{i=1}^m \frac{1}{\bar{n} \cdot q_i} \,.
\end{eqnarray}
The factor of $1/m!$ is from the presence of $m$ identical $A_c$ fields at the
same point.  Thus, we have the tree level result
\begin{eqnarray} \label{J2}
  J^{\rm eft}_{\rm had}\bigg|_{\mu=m_b} &=& \bar\xi_{n,p}\: 
 \exp\bigg( \frac{g\,\bar{n}\cdot A_{n,q}}{\bar{n}\cdot q} \bigg) 
 \:\Gamma\, h_v \,.
\end{eqnarray}
We can rewrite the exponential in the above expression using the 
projection operator ${\cal P}$:
\begin{eqnarray} \label{J2b}
  \exp\bigg( \frac{g\,\bar{n}\cdot A_{n,q}}{\bar{n}\cdot q} \bigg) &=& 
  \sum_{{\rm perms}} \exp\bigg( \frac{g\,\bar{n}\cdot A_{n,q}}{\bar{{\cal P}}} \bigg) 
  \equiv W^\dagger \,.
\end{eqnarray}
Where we have given the above object a name since it will occur again and again. 
Why is $W$ so important? Consider a collinear gauge transformation on the current
in Eq.~(\ref{J2}).  The field $h_v$ is invariant since collinear gluons do not couple to 
heavy quarks,  on the other hand, the collinear quark field transforms as 
$\xi_{n,p} \to e^{i\alpha(x)} \xi_{n,p}$.  Thus, the operator $\bar\xi_{n,p} \Gamma h_v$ 
is not gauge invariant. However, we find that 
\begin{eqnarray}
 &&\exp\bigg(\frac{g\,\bar{n}\cdot A_{n,q}}{\bar{n}\cdot q} \bigg)  \to 
   \exp\bigg(\frac{g\,\bar{n}\cdot A_{n,q}}{\bar{n}\cdot q} \bigg)
   \exp\Big[ i\alpha(x) \Big] \,,
\end{eqnarray}
and the last exponential exactly cancels the transformation of $\bar\xi_{n,p}$.
By gauge invariance the current therefore has to be of the form in
Eq.~(\ref{J2}) for an arbitrary scale $\mu$.  It is convenient to define a field
that transforms as a singlet under a collinear gauge transformation
\begin{eqnarray}
  \chi_{n} = W  \xi_{n,p}\,.
\end{eqnarray}
We will refer to $\chi_{n}$ as the jet field since it involves a collinear
quark field plus an arbitrary number of collinear gluons moving in the $n$
direction. 

Hard fluctuations in the full theory do not occur in the effective theory since 
they have been integrated out. However, the physics of the hard fluctuations
appears in the Wilson coefficients of the effective theory as a result of matching. 
The SCET Wilson coefficients can therefore be nontrivial functions of the large 
collinear momentum, $C(\bar{n}\cdot p_i)$. Fortunately collinear gauge invariance 
restricts these coefficients so that they only depend on the linear combination picked out
by $\bar{{\cal P}}$. Thus the general Wilson coefficient in SCET will be a function
$C(\bar{{\cal P}}, \bar{{\cal P}}^\dagger)$ which must be inserted between gauge
invariant products of collinear fields and $W$.
Thus in terms of the $\chi_n$ field the
leading order effective theory current for $Q\lambda < \mu < m_b$ has the form
\begin{eqnarray}   \label{eftJ}
 J^{\rm eft}_{\rm had} &=& \bar\chi_{n}\: C_i(\mu,\bar{{\cal P}}^\dagger) \: 
 \Gamma\, h_v \,,
\end{eqnarray}
where $\bar{{\cal P}}^\dagger$ acts to the left.
For a non-abelian gauge group a similar gauge invariance argument applies,
however the matching in Fig.~\ref{fig_current} is more
complicated. In momentum space we find
\begin{eqnarray} \label{jetq}
 \chi_{n} &=& \sum_k \sum_{\rm perms}
  \frac{(-g)^k}{k!}\left( {\bar n\cdot A_{\bar{n}, q_1}\cdots
  \bar n\cdot A_{\bar{n}, q_k} \over
  [\bar{n}\cdot q_1] [\bar{n}\cdot (q_1+q_2)]\cdots[\bar{n}\cdot \sum_{i=1}^k q_i] }
  \right) \xi_{n,p}
  \nonumber \\
   &=& \sum_{{\rm perms}} \exp\bigg( \frac{-g\,\bar{n}\cdot A_{n,q}}{\bar{{\cal P}}} \bigg)  \xi_{n,p}
   = W \xi_{n,p} \,.
\end{eqnarray}

\section{Factorization}

A remarkable consequence of the gauge symmetries of SCET is the factorization of 
usoft and collinear effects. Towards this end we introduce the usoft Wilson line
\begin{equation}
Y(x) = {\rm P exp} \bigg( ig \int^x_{-\infty} ds \; n\cdot A_{us}(ns) \bigg) \,,
\end{equation}
and redefine the collinear fields as follows:
\begin{equation}\label{fieldred}
\xi_{n,p} = Y \xi^{(0)}_{n,p}   \hspace{5ex} A^\mu_{n,q} = Y A^{(0) \mu}_{n,q} Y^\dagger \,.
\end{equation}
Under these field redefinitions the usoft gluons decouple from the collinear fields. In 
other words 
the leading Lagrangian in Eq.~(\ref{scetlag}) becomes completely independent of the usoft fields. 
At higher orders in the $\lambda$ expansion this decoupling does not occur. 
Furthermore as we will see decoupling of usoft gluons in the leading 
Lagrangian does not necessarily mean that usoft gluons decouple 
in subleading operators or currents.   
 
Let us now return to our example and study the consequences of the above field redefinition.
The inclusive photon energy spectrum can be written using the optical theorem as
\begin{eqnarray}
 \frac{1}{\Gamma_0(m_b)}\frac{d\Gamma}{dE_\gamma} = 
 \frac{4 E_\gamma}{m_b^3} \left(-\frac{1}{\pi} \right)
 {\mbox{Im }} T(E_\gamma)\,,
\end{eqnarray}
where the forward scattering amplitude $T(E_\gamma)$ is 
\begin{eqnarray} \label{Tg}
  T(E_\gamma) = \frac{i}{m_B}\int\!\! d^4x\, e^{-iq\cdot x}\, \langle \bar B| 
  T J_\mu^\dagger(x) J^\mu(0)|\bar B\rangle \,,
\end{eqnarray}
with relativistic normalization for the $|\bar B\rangle$ states.  The
current is $J_\mu = \bar s\,i\sigma_{\mu\nu} q^\nu P_R\, b$, and
$\Gamma_0(m_b)$ is given in Eq.~(\ref{gamma0})

In the endpoint region  we match the current in the 
time ordered product in Eq.~(\ref{Tg}) onto SCET fields. 
At leading order in $\lambda$ 
\begin{eqnarray}\label{Jeff}
 J_\mu &=& -E_\gamma\,  e^{i( \bar{{\cal P}} \frac{n}{2}+{\cal P}_\perp - m_b v)\cdot x }\,
    \bigg\{ \big[2 C_9(\bar{{\cal P}},\mu)+C_{12}(\bar{{\cal P}},\mu)\big] J^{\rm eff}_\mu 
  -  C_{10}(\bar{{\cal P}}, \mu) {\widetilde J}^{\rm eff}_\mu \bigg\}  \,, 
\end{eqnarray}
where
\begin{eqnarray} \label{Jeff3}
 J^{\rm eff}_\mu &=&  \bar \xi_{n,p}\, W \gamma_\mu^\perp P_L \, b_v \,,\qquad
 {\widetilde J}^{\rm eff}_\mu = \bar{n}_\mu \bar\xi_{n,p}\,  W P_R\, b_v\,. 
\end{eqnarray}
The SCET Wilson coefficients $C_{9,10,12}(\bar{{\cal P}},\mu)$ are given at one-loop in
Ref.~\cite{Bauer:2000yr}. In Eq.~(\ref{Jeff}) label conservation 
sets $\bar{{\cal P}}=m_b$ and ${\cal P}_\perp=0$.  The
current $\tilde J^{\rm eff}_\mu$ does not contribute for real transversely
polarized photons so I drop it. Inserting
Eq.~(\ref{Jeff}) into Eq.~(\ref{Tg}) gives
\begin{eqnarray}
   \frac{4E_\gamma}{m_b^3}\: T(E_\gamma) 
    \equiv H(m_b,\mu)\: T^{\rm eff}(E_\gamma,\mu) \,,
\end{eqnarray}
where $T^{\rm eff}$ is the forward scattering amplitude in the effective theory
\begin{eqnarray}
  T^{\rm eff} &=& i \int\! d^4x\: e^{i(m_b\frac{\bar{n}}{2} - q )
  \cdot x} \: \Big\langle \bar B_v \Big|\,{\rm T}\, J_\mu^{\rm eff\dagger}(x)\, 
   J^{\rm eff \mu}(0)\, \Big| \bar B_v \Big\rangle \,,
\end{eqnarray}
with HQET normalization for the states.  The hard amplitude is
\begin{eqnarray} \label{H}
  H(m_b, \mu) = \frac{16 E_\gamma^3}{m_b^3}\:  
  \Big| C_9(m_b,\mu) + \frac12 C_{12}(m_b,\mu) \Big|^2  \,.
\end{eqnarray}

Next the usoft gluons are decoupled from the collinear
fields by the field redefinitions in Eq.~(\ref{fieldred}) along with
\begin{eqnarray}
   W \to Y W^{(0)} Y^\dagger\,,
\end{eqnarray}
which is a consequence of Eq.~(\ref{fieldred}).
This gives  
\begin{eqnarray} \label{Jeff2}
 J^{\rm eff}_\mu = \bar \xi^{(0)}_{n,p}\, W^{(0)} \gamma_\mu^\perp P_L \, 
  Y^\dagger \, b_v \,,
\end{eqnarray}
Thus, the time-ordered product of the effective theory currents is
\begin{eqnarray}\label{Tfact1}
 T^{\rm eff} &=& i \int\! d^4x\: e^{i(m_b \frac{\bar{n}}{2} - q )\cdot x} 
  \: \Big\langle \bar B_v \Big|\, {\rm T}\, 
  \big[ \bar b_v Y \big](x)\: \big[ Y^\dagger b_v \big](0)\, 
  \Big|\bar B_v \Big\rangle \\
& & \hspace{5ex} \times
 \Big\langle 0 \Big| \,\mbox{T}\, \big[ W^{(0)\dagger} \xi_{n,p}^{(0)}\big](x)\: 
 \big[ \bar\xi_{n,p}^{(0)} W^{(0)}\big] (0)\, \Big| 0 \Big\rangle \nonumber \,.
\end{eqnarray}
In the effective theory the Hilbert space of states is the direct product of
the usoft and collinear Hilbert states. As a consequence the $\bar B$ meson state 
contains no collinear particles, and the collinear physics can be separated from the
usoft physics. Next introduce the Fourier transform 
\begin{eqnarray}\label{SCdef}
 \Big\langle 0 \Big| \,\mbox{T}\, \big[ W^{(0)\dagger} \xi_{n,p}^{(0)}\big](x)\: 
 \big[ \bar\xi_{n,p}^{(0)} W^{(0)}\big] (0)\, \Big| 0 \Big\rangle \equiv
 i\, \int\!\frac{d^4 k}{(2\pi)^4}\, e^{-ik\cdot x}\, J(P,k) 
 \: \frac{{n\!\!\!\slash}}{2} \,,
\end{eqnarray}
where $P$ is sum of the label momenta carried by the collinear
fields. By momentum conservation $P = m_b$.
Because the collinear Lagrangian contains only the $n\cdot \partial$ derivative $J(p,k)$ 
only depends on the component $k^+ = n\cdot k$ of the residual momentum $k$. 
This allows us to perform the $k_-, k_\perp$ integrations in Eq.~(\ref{SCdef}) which puts $x$ on
the light cone
\begin{eqnarray} \label{Tfact2}
 T^{\rm eff} &=& \frac12 \int\! d^4 x\,e^{i(m_b \frac{\bar{n}}{2}-q)\cdot x}\: 
 \delta(x^+) \delta({\vec x}_\perp) \int\!\frac{d k^+}{2\pi}\, 
 e^{-\frac{i}{2} k^+ x^-}\, \Big\langle \bar B_v \Big|\, {\rm T}\, 
 \big[ \bar b_v Y \big](x)\: \big[ Y^\dagger b_v \big](0)\, 
 \Big|\bar B_v \Big\rangle \: J(P,k^+) \nonumber\\ 
&=& \frac12 \int\! {d k^+}\,  J(P,k^+)\,\int\! \frac{dx^-}{4\pi}\, 
 e^{-\frac{i}{2}(2 E_\gamma-m_b+k^+) x^-}\, \Big\langle \bar B_v \Big| \, 
 {\rm T}\, \big[ \bar b_v Y \big](\mbox{\small $\frac{n}{2}$}\,x^-)\: \big[ 
  Y^\dagger b_v \big](0)\, \Big|\bar B_v \Big\rangle  \,.  
\end{eqnarray}
The typical offshellness of the collinear particles is $p^2\sim m_b
\Lambda_{\rm QCD}$ so the function $J(P,k^+)$ can be calculated perturbatively.  At
lowest order in $\alpha_s(\mbox{\small $\sqrt{m_b\Lambda_{\rm QCD}}\,$})$,
$J(P,k^+)$ is determined by the collinear quark propagator carrying momentum
$(P+k)$
\begin{eqnarray}
  J(P,k^+) = \frac{\bar{n}\cdot P}{(P+k)^2 + i\epsilon} = 
   \frac{1}{n\cdot k + P_\perp^2/(\bar{n}\cdot P) + i\epsilon}\,.
\end{eqnarray}
The remaining matrix element in Eq.~(\ref{Tfact2}) is purely usoft
\begin{eqnarray}\label{Sdef}
 S(l^+) &\equiv & \frac12 \int \frac{dx^-}{4\pi}\, e^{-\frac{i}{2}\, l^+ x^-}
 \Big\langle \bar B_v \Big| \,T\,\big[\bar b_v Y\big](\mbox{\small $\frac{n}{2}$}
 \, x^- )\: \big[Y^\dagger b_v\big](0)\,\Big| \bar B_v \Big\rangle \,.
 \end{eqnarray} 
Inserting this into Eq.~(\ref{Tfact2}) and taking the imaginary part
gives
\begin{eqnarray}\label{Tfact3}
 \frac{1}{\Gamma_0} \frac{d\Gamma}{dE_\gamma}
 = H(m_b, \mu)\int_{2E_\gamma-m_b}^{\bar \Lambda} \!\!\!\! dk^+\: S(k^+, \mu)\:
   J(k^+ + m_b - 2E_\gamma , \mu)\,.
\end{eqnarray}
where
\begin{eqnarray}\label{Jdef}
  J(k^+) \equiv -\frac{1}{\pi}\,\mbox{Im}\: J(P,k^+)\,.
\end{eqnarray}
This result  is valid to all orders in $\alpha_s$ and leading order 
in $\Lambda_{\rm QCD}/Q$ where $Q=E_\gamma$ or $m_b$.  
The individual terms in Eq.~(\ref{Tfact3}) depend  on the 
scale $\mu$ in such a way that the decay rate is $\mu$ independent. In the next
section I discuss the $\mu$ dependence of $S$ and $J$ in detail.

\section{Renormalization Group: Summing Logarithms}

Eq.~(\ref{qcdinclusive}) gives the tree level and $\alpha_s$ corrections to the 
$B \to X_s + \gamma$ decay rate. The leading order contribution is proportional 
to $\delta(1-x)$, and   the next-to-leading order corrections have
contributions of the form $\alpha_s \ln(1-x)/(1-x)$, where $x = 2 E_\gamma/m_b$.  
Clearly  when $x \sim 1-\alpha_s$ 
these corrections are large and must be resummed. This can be
accomplished in a straightforward manner by using the renormalization group equations 
of SCET. The resummation is most easily carried out by taking
moments  with respect to $x$, then the large corrections as $x\rightarrow 1$ become 
large logs of $N$ in the expression for the $N$th moment. 
  
Taking moments of the factored decay rate  in Eq.~(\ref{Tfact3}) gives
\begin{eqnarray}\label{moment}
\int_0^1dx \;  x^N \frac{1}{\Gamma_0} \frac{d\Gamma}{dx} &=&\tilde{H}(m_b,\mu) 
\int_0^1dx\,x^N \int_x^1 d\xi \,S(\xi,\mu) \, J(m_b(\xi-x),\mu) \\
&=&\tilde{H}(m_b,\mu) \int_0^1 d\xi \, \xi^{N+1}\int_0^1 du \,u^N \,S(\xi,\mu) \, J(m_b\xi(1-u),\mu) \,,
\nonumber 
\end{eqnarray}
where $\xi = k^+/m_b + 1$.
In the last line we made the substitution $x = u \, \xi$. 
Since the large logs come from the region $\xi, x \sim1$, 
the factor of $\xi$ in the argument of the jet function can be set equal to 1. 
Then the moments factor:
\begin{eqnarray}\label{prod}
\Gamma_N = H(m_b,\mu) \, S_N(\mu) \, J_N(\mu) \, ,
\end{eqnarray}
where  
\begin{eqnarray}\label{moments}
\Gamma_N &=& \int_0^1dx \, x^N \, \frac{1}{\Gamma_0} \frac{d\Gamma}{dx}   \, , \\
 S_N &=& \int_0^1d\xi \, \xi^N \, S(\xi,\mu) \, , \nonumber \\
 J_N(\mu) &=& \int_0^1 du \,u^N \,J(m_b(1-u),\mu) \nonumber \, .
\end{eqnarray}
We are only interested in the large $N$ moments, so have used $S_{N+1} = S_N + {\cal O}(1/N)$.

\begin{figure}
\begin{center}
\includegraphics[width=6in]{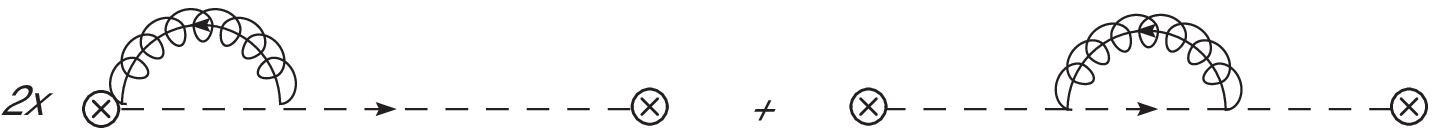}
\caption{\it Graphs needed  to calculate the ${\cal O}(\alpha_s)$ counterterm to $J_N$. 
\label{figJF}}
\end{center}
\end{figure}

\begin{figure}
\begin{center}
\includegraphics[width=4in]{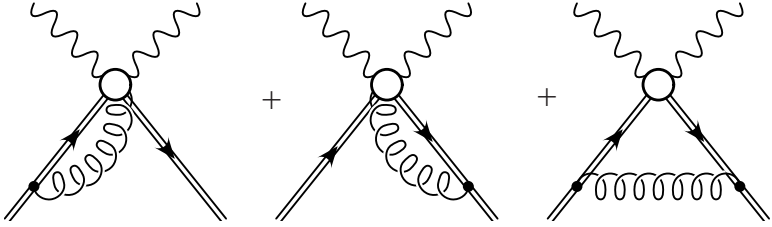}
\caption{\it Graphs needed  to calculate the ${\cal O}(\alpha_s)$ counterterm to $S_N$. 
\label{figSF}}
\end{center}
\end{figure}

To resum logarithms we use the renormalization group equations for $S_N$ and $J_N$ 
The one loop anomalous dimension for the jet function renormalization group equations 
is calculated from  the diagrams in Fig.~(\ref{figJF}). The result is
\begin{eqnarray}\label{jrge}
\mu \frac{d}{d\mu} J_N(\mu) = \left[ \frac{2 C_F \alpha_s}{\pi}\log\left( \frac{\bar \mu^2}
{m^2_b} \frac{N}{N_0}\right) +\frac{3\alpha_s}{2\pi} C_F
\right] J_N(\mu) \, ,
\end{eqnarray}
where $N_0=e^{-\gamma}$. 
The one loop anomalous dimension of $S_N$ is determined from the diagrams in Fig~(\ref{figSF}), and
the renormalization group equation immediately follows: 
\begin{eqnarray}\label{srge}
\mu \frac{d}{d\mu} S_N(\mu) = \left[ -\frac{2 C_F \alpha_s}{\pi}\log\left( \frac{\bar \mu}{m_b}
 \frac{N}{N_0}\right) + \frac{\alpha_s C_F}{\pi} \right] S_N(\mu)\, .
\end{eqnarray}


Defining  $y_0 = (N_0/N)$  logarithms in the hard, jet and usoft
functions are minimized at the scales $m_b, m_b \sqrt{y_0}$ and $m_b y_0$ respectively. 
Large logarithms of $N$  are summed by evolving the jet and usoft functions to these scales. 
The evolution can also be done in one step by defining separate renormalization scales for collinear 
and usoft loops~\cite{Fleming:2003gt}. 
Loops whose momenta scale like $(1,\lambda^2, \lambda)$ come with a factor of $\mu_c^{4-D}$ 
and loops whose momenta scale like $(\lambda^2,\lambda^2,\lambda^2)$ come with a factor 
$\mu_u^{4-D}$. This idea is similar to the velocity renormalization group in
NRQCD \cite{Luke:1999kz}.  The renormalization group equations for $J_N$ and $S_N$ take the form
\begin{eqnarray}\label{twomu}
\mu_c \frac{d} {d \mu_c} J_N &=& \gamma_J^N(\mu_c) J_N \, ,\\
\mu_u \frac{d} {d \mu_u} S_N &=& \gamma_S^N(\mu_u) S_N \, .\nonumber
\end{eqnarray}
Factorization of usoft and collinear degrees of freedom guarantees that $\gamma_J$ is a function  of $\mu_c$ only and
that $\gamma_S$ is a function of $\mu_u$ only. The scales are however correlated,  so that $\mu_c = m_b\sqrt{y}$ and
$\mu_u = m_b y$. Evolving the variable $y$ from $1$ to $y_0$ simultaneously resums large logs in both $J_N$
and $S_N$. 

Defining $\tilde{\Gamma}_N =J_N S_N$, the evolution equation for $\tilde{\Gamma}_N$ as function of $y$ is
\begin{eqnarray}\label{yrge}
y \frac{d} {d y} \tilde{\Gamma}_N = \left(\frac{1}{2}\gamma_J^N(m_b \sqrt{y}) +\gamma_S^N(m_b y)\right)
\tilde{\Gamma}_N \, .
\end{eqnarray}
This equation is easily integrated to obtain  the following expression for the resummed moments:
\begin{equation}
\label{fullyresummed}
\Gamma_N = 
H(m_b) S_N(m_b y_0)\, e^{\log(N) g_1(\chi) + g_2(\chi)},
\end{equation}
where
\begin{eqnarray}
\label{gis}
g_1(\chi) &=&
-\frac{2 C_F}{\beta_0\chi}\left[(1-2\chi)\log(1-2\chi)
-2(1-\chi)\log(1-\chi)\right], \\
g_2(\chi) &=& -\frac{8 \Gamma_2}{\beta_0^2}
  \left[-\log(1-2\chi)+2\log(1-\chi)\right] - \log(1-\chi)\nonumber\\
 && - \frac{2C_F\beta_1}{\beta_0^3}
   \left[\log(1-2\chi)-2\log(1-\chi)
  +\frac12\log^2(1-2\chi)-\log^2(1-\chi)\right] \nonumber\\
 && -
 \frac{2C_F}{\beta_0} \log(1-2\chi) -
\frac{4C_F}{\beta_0}\log (N_0) \left[\log(1-2\chi)-\log(1-\chi)\right]  -
 \frac{3C_F}{\beta_0} \log(1-\chi) \,,\nonumber
\end{eqnarray}
where
$\chi=\log (N)\, \alpha_s(m_b)\beta_0/4\pi$, $\Gamma_2 = C_F[C_A(67/36 - \pi^2/12) - 5n_f/18]$, 
$\beta_0 = (11C_A-2n_f)/3$, and
$\beta_1 = (34C_A^2-10C_A n_f-6C_F n_f)/3$.  $\Gamma_2$ is 
the $O(\alpha_s^2)$ piece of the cusp anomalous dimension, which was taken from Ref.~\cite{Korchemsky:1985ts,Korchemsky:1992xv}.

The expression in Eq.~(\ref{fullyresummed}) gives the resummed expression for 
the moments of the differential cross section to next-to-leading logarithmic
order. To obtain the differential cross section, the inverse-Mellin transform  of 
Eq.~(\ref{fullyresummed}) must be taken.  Using the results of
Ref.~\cite{Leibovich:1999xf}, we find:
\begin{eqnarray}
\label{zspaceresummed}
\frac{d\Gamma}{dx}&=& -\int_x^1 \frac{d\xi}{\xi}\, \Gamma_0(m_b)\,S\left(\xi\right)\\
&&\phantom{ -\int_x^1 d\xi\, }
x \frac{d}{dx} \left\{
\theta(\xi-x) \; \frac{\exp [ l g_1(\alpha_s \beta_0 l/(4\pi)) +
g_2(\alpha_s \beta_0 l/(4\pi))]}{\Gamma[1-g_1(\alpha_s \beta_0
l/(4\pi)) - \alpha_s \beta_0 l/(4\pi) g_1^\prime(\alpha_s \beta_0
l/(4\pi))]}\right\} \,,\nonumber
\end{eqnarray}
where $l \approx -\log(\xi-x)$, $\alpha_s \equiv \alpha_s(m_b)$, and the shape function $S$ 
contains no large logarithms. 

\section{Conclusion}

By introducing soft collinear effective theory within the context of
$B \to X_s \gamma$ decay near the endpoint I hope I have been able 
to shed some light on some of the developments that have taken place in  
this field recently. There has been much work that I have not been able 
to cover. In particular I have not discussed the application of SCET to exclusive
decays. For an up-to-date discussion of recent  progress on exclusive rare
and semileptonic $B$ decays I refer you to Dan Pirjol's talk~\cite{Pirjol:2003ef}. 

\section*{Acknowledgments}

I would like to thank my collaborators on the various projects that have been incorporated
into this talk: Christian Bauer, Adam Leibovich, Mike Luke, Tom Mehen, Dan Pirjol, and
Iain Stewart. This work was supported in part by Department of Energy grant DOE-ER-40682-143. 
The preprint number for this report is CMU-HEP-03-11.

%%%%%%%%%%%%%%%%%%%%%%%%%%%%%%%%%%%%%
% Label to flag the last page of your contribution
% Replace Perret by your name starting with a capital letter
%
\label{FlemingEnd}
 
\end{document}

%% file: fpcp03-macro.tex
\def\beq{\begin{equation}}
\def\eeq#1{\label{#1}\end{equation}}
\def\eeqn{\end{equation}}

\def\beqa{\begin{eqnarray}}
\def\eeqa#1{\label{#1}\end{eqnarray}}
\def\eeqan{\end{eqnarray}}

\def\overbar#1{\overline{#1}}

\let\bar=\overbar

\def\cD{{\cal D}}

\def\Dslash{\not{\hbox{\kern-4pt $D$}}}
\def\dslash{\not{\hbox{\kern-2pt $\del$}}}

\def\msb{{\bar{\ssstyle M \kern -1pt S}}}

\def\BB0bar{B^0 {\overline B}^0}
\def\BB0dbar{B_d^0 {\overline B}_d^0}
\def\BB0sbar{B_s^0 {\overline B}_s^0}

\RequirePackage{xspace}

\usepackage{relsize}
\def\babar{\mbox{\slshape B\kern-0.1em{\smaller A}\kern-0.1em
    B\kern-0.1em{\smaller A\kern-0.2em R}}}

\def\Kbar  {\kern 0.2em\overline{\kern -0.2em K}{}\xspace}

\def\Kz    {\ensuremath{K^0}\xspace}
\def\Kzb   {\ensuremath{\Kbar^0}\xspace}
\def\KzKzb {\ensuremath{\Kz \kern -0.16em \Kzb}\xspace}
\def\Kp    {\ensuremath{K^+}\xspace}
\def\Km    {\ensuremath{K^-}\xspace}

\def\KpKm  {\ensuremath{\Kp \kern -0.16em \Km}\xspace}

\def\Dbar    {\kern 0.2em\overline{\kern -0.2em D}{}\xspace}

\def\Dz      {\ensuremath{D^0}\xspace}
\def\Dzb     {\ensuremath{\Dbar^0}\xspace}
\def\DzDzb   {\ensuremath{\Dz {\kern -0.16em \Dzb}}\xspace}
\def\Dp      {\ensuremath{D^+}\xspace}
\def\Dm      {\ensuremath{D^-}\xspace}

\def\DpDm    {\ensuremath{\Dp {\kern -0.16em \Dm}}\xspace}

\def\Bbar    {\kern 0.18em\overline{\kern -0.18em B}{}\xspace}

\def\BB      {\ensuremath{B\Bbar}\xspace} 
\def\Bz      {\ensuremath{B^0}\xspace}
\def\Bzb     {\ensuremath{\Bbar^0}\xspace}
\def\BzBzb   {\ensuremath{\Bz {\kern -0.16em \Bzb}}\xspace}
\def\Bu      {\ensuremath{B^+}\xspace}
\def\Bub     {\ensuremath{B^-}\xspace}

\def\BpBm    {\ensuremath{\Bu {\kern -0.16em \Bub}}\xspace}

\mathchardef\Upsilon="7107
\def\Y#1S{\ensuremath{\Upsilon{(#1S)}}\xspace}% no space before {...}!

\mathchardef\Deltares="7101
\mathchardef\Xi="7104
\mathchardef\Lambda="7103
\mathchardef\Sigma="7106
\mathchardef\Omega="710A

\def\Deltabar{\kern 0.25em\overline{\kern -0.25em \Deltares}{}\xspace}
\def\Lbar{\kern 0.2em\overline{\kern -0.2em\Lambda\kern 0.05em}\kern-0.05em{}\xspace}
\def\Sigbar{\kern 0.2em\overline{\kern -0.2em \Sigma}{}\xspace}
\def\Xibar{\kern 0.2em\overline{\kern -0.2em \Xi}{}\xspace}
\def\Obar{\kern 0.2em\overline{\kern -0.2em \Omega}{}\xspace}
\def\Nbar{\kern 0.2em\overline{\kern -0.2em N}{}\xspace}
\def\Xb{\kern 0.2em\overline{\kern -0.2em X}{}\xspace}

\newcommand{\tev}{\ensuremath{\mathrm{\,Te\kern -0.1em V}}\xspace}
\newcommand{\gev}{\ensuremath{\mathrm{\,Ge\kern -0.1em V}}\xspace}
\newcommand{\mev}{\ensuremath{\mathrm{\,Me\kern -0.1em V}}\xspace}
\newcommand{\kev}{\ensuremath{\mathrm{\,ke\kern -0.1em V}}\xspace}
\newcommand{\ev}{\ensuremath{\mathrm{\,e\kern -0.1em V}}\xspace}
\newcommand{\gevc}{\ensuremath{{\mathrm{\,Ge\kern -0.1em V\!/}c}}\xspace}
\newcommand{\mevc}{\ensuremath{{\mathrm{\,Me\kern -0.1em V\!/}c}}\xspace}
\newcommand{\gevcc}{\ensuremath{{\mathrm{\,Ge\kern -0.1em V\!/}c^2}}\xspace}
\newcommand{\mevcc}{\ensuremath{{\mathrm{\,Me\kern -0.1em V\!/}c^2}}\xspace}
\def\mus  {\ensuremath{\rm \,\mus}\xspace}

\def\mus        {\ensuremath{\,\mu{\rm s}}\xspace}    %% microsecond
\def\to                 {\ensuremath{\rightarrow}\xspace}

\def\pep2{PEP-II}

\def\gsim{{~\raise.15em\hbox{$>$}\kern-.85em
          \lower.35em\hbox{$\sim$}~}\xspace}
\def\lsim{{~\raise.15em\hbox{$<$}\kern-.85em
          \lower.35em\hbox{$\sim$}~}\xspace}

\newcommand{\as}{\ensuremath{\alpha_{\scriptscriptstyle S}}\xspace}

\xspace

\newcommand{\lqcd}{\ensuremath{\Lambda_{\mathrm{QCD}}}\xspace}

\def\jetset74   {\mbox{\tt Jetset \hspace{-0.5em}7.\hspace{-0.2em}4}\xspace}